\documentclass[pra,twocolumn]{revtex4}
\usepackage{amsmath}
\usepackage{amsfonts}
\usepackage{graphicx}
\usepackage{epsfig}

\begin{document}

\title{Stability of gap solitons in a Bose-Einstein condensate}
\author{Karen Marie Hilligs{\o}e$^{1,2}$}
\author{Markus K. Oberthaler$^1$}
\author{Karl-Peter Marzlin$^1$}
\affiliation{$^1$Fachbereich Physik der Universit\"at Konstanz, Fach M674,
    D-78457 Konstanz, Germany}
\affiliation{$^2$Department of Physics and Astronomy, University of Aarhus, DK-8000 \AA rhus C, Denmark }
\date{\today}

\begin{abstract}
We analyze the dynamical stability of gap solitons formed in a 
quasi one-dimensional
Bose-Einstein condensate in an optical lattice. Using two different
numerical methods we show that, under realistic assumptions for 
experimental parameters, a gap soliton is stable only in a truly
one-dimensional situation. In two and three dimensions resonant
transverse excitations lead to dynamical instability.
The time scale of the decay is numerically calculated and shown to
be large compared to the characteristic time scale of solitons
for realistic physical parameters.
\end{abstract}
\maketitle
\section{Introduction}
One of the most fundamental facts about Bose-Einstein condensates
(BEC) of dilute atomic gases is that they can be very well described
by the Gross-Pitaevskii equation (GPE)
\begin{equation}
  i\hbar \dot{\psi}({\bf x},t) = \left ( \frac{{\bf p}^2}{2M} + V({\bf x}) +
  \kappa |\psi({\bf x},t)|^2 \right ) \psi({\bf x},t)
\label{gpe}
\end{equation}
where $\psi$ denotes the collective wavefunction of condensed atoms
with mass $M$ and coupling constant 
$\kappa := 4\pi\hbar^2 a N_A/M$, 
with $a$ being the s-wave scattering length and 
$N_A$ denoting the number of atoms in the BEC
(see, e.g., Ref.~\cite{castin01}). 
Therefore a BEC provides a physical realization of many nonlinear wave
phenomena among which the formation of solitons is particularly
interesting. 

For our purposes solitons 
are wavepackets in which the dispersive effect
of the kinetic term is exactly cancelled by the nonlinear interaction
energy so that their shape does not change. Two fundamental types of
solitons have so far been experimentally realized in a BEC: dark
solitons \cite{sengstock99,denschlag00} correspond to a stable density
dip in a BEC of repulsive atoms.
Bright solitons do exist for atoms with attractive interaction
($\kappa <0$) and are described by a one-dimensional solution of the
GPE for vanishing potential $V({\bf x})$,
\begin{equation} 
  \psi_{\mbox{{\scriptsize bright}}}(z,t) = 
  \frac{1}{\sqrt{2\textrm{w}}} \mbox{sech}(z/\textrm{w})e^{-i\omega_s t},
\label{brightsol}
\end{equation} 
where $\textrm{w}=2\hbar^2/(M\kappa_{1D})$ and 
$\omega_s=\hbar/(2M\textrm{w}^2)$. To relate the
one-dimensional GPE to its three-dimensional origin we assumed that
the BEC is tightly trapped in the transverse direction so that
no transverse excitations can occur. Taking the BEC to be in the
transverse ground state amounts to replacing the three-dimensional
coupling constant $\kappa$ by
$\kappa_{1D}=\kappa/A_{\perp}$, where $A_{\perp}=2\pi a_\perp^2$ 
with $a_\perp := \sqrt{\hbar/(M \omega_\perp)}$ is the 
harmonic oscillator length of the transverse potential.
The attraction between the atoms prevents the dispersion of the
sech-shaped wavepacket. Note that for $\kappa_{1D} > 0$,
corresponding to repulsive atom-atom interaction,
the kinetic energy and the interaction energy 
have the same sign and therefore cannot cancel each other, a bright
soliton is then not possible.
Very recently bright solitons have been created in a quasi
one-dimensional setup \cite{salomon02,hulet02} 
where the transverse potential $V(x,y)$ tightly confines the
BEC, thus suppressing transverse excitations and three-dimensional collapse.

In this paper we are concerned with the dynamical stability of gap
solitons. This collective state, which has not yet been realized
experimentally, exists for repulsive atoms ($\kappa_{1D} >0$) 
in a periodic potential $V(z)$ and is related to bright solitons. 
The basic idea of a gap soliton is the following: 
as is well known the energy eigenvalues of noninteracting particles
in a periodic potential are given by
energy bands $E_n(q)$, where $n$ is the band index and $q$ denotes
the quasi momentum. If a particle's state is prepared in the lowest
band only the dispersion relation $p^2/(2M)$ in free space is replaced
by the lowest band energy $E_0(q)$. 
Around the upper band edge, which we take
to be at $q=0$, this energy
can be approximated by $E_0(q) \approx E_0(0)
+ q^2 /(2 M^*)$ where $M^* := (d^2 E_0(q)/dq^2)^{-1}|_{q=0}$ is the
effective mass of the particle in the periodic potential. Since at the
upper band edge $M^* <0$ the ``kinetic energy'' becomes negative and
a cancellation with the positive interaction energy becomes possible.
The corresponding state is called a {\em gap soliton}.

Gap solitons have been realized in nonlinear optics using a periodic 
modulation of the propagation medium \cite{chen87}. In 
nonlinear atom optics they have first been predicted by Lenz {\em et al.} 
\cite{lenz94} in the context of 
light-induced nonlinearities \cite{krutitsky99,krutitsky02}.
Here we are concerned with the collision-induced nonlinearity
appearing in Eq.~(\ref{gpe}). 

We consider a BEC that is placed in a
one-dimensional optical lattice, created by
far-detuned laser light with wavenumber $k_L$,
which produces an optical potential of the form  
$V(z) = - V_0 \cos (2 k_L z)$, and is subject
to a tight harmonic transverse confinement of the form
$V_\perp (x,y) = M \omega_\perp^2 (x^2+y^2)/2$. 
Here $\omega_\perp$ is the transverse trap frequency and
$V_0$ the strength of the optical potential.
The derivation of the corresponding gap soliton solution of the 
one-dimensional GPE is tedious
and includes a multiple scales analysis. For nonlinear optics
it has been derived by Sipe {\em et al.} \cite{sipe88}. For 
nonlinear atom optics a related  derivation has been sketched in 
Refs.~\cite{lenz94,steel98,kon01}. One finds that,
within the effective mass approximation, 
the one-dimensional gap soliton is described by
\begin{equation} 
  \psi_{\mbox{{\scriptsize gap}}}(z,t) 
   = \varphi_{\mbox{{\scriptsize be}}}(z,t)                
               \frac{1}{\sqrt{2\tilde{\textrm{w}}}} 
               \mbox{sech}(z/\tilde{\textrm{w}})
               e^{-i\tilde{\omega}_s t},
\label{gapsol}
\end{equation} 
where $\varphi_{\mbox{{\scriptsize be}}}(z,t)=
\varphi_{\mbox{{\scriptsize be}}}(z) e^{-i E_0(0) t/\hbar}$ is the
solution of the linear Schr\"odinger equation that corresponds to
the upper band edge. The parameters 
$\tilde{\textrm{w}}$ and $\tilde{\omega}_s$
have the same form as $\textrm{w}$ and $\omega$ for the bright soliton but
with $M$ replaced by $|M^*|$ and $\kappa_{1D}$ replaced by $\tilde{\kappa}_{1D}
:= \kappa_{1D} \int |\varphi_{\mbox{{\scriptsize be}}}(z)|^4 dz$.
Apart from $\varphi_{\mbox{{\scriptsize be}}}$
the solution (\ref{gapsol}) just corresponds to a bright soliton
for a particle of mass $|M^*|$ and coupling constant $\tilde{\kappa}$.

The range of experimentally promising values for the optical potential 
strength $V_0$, the number of condensed atoms $N_A$,
and the transverse confinement frequency $\omega_\perp$
has been examined in a study by Brezger {\em et al.}
\cite{brezger02} following standard textbooks \cite{agrawal89}. 
It was found that values around $N_A = 400$,
$V_0 = \hbar^2 k_L^2/(2M)$ 
and $\omega_\perp = 110$ s$^{-1}$ should
be optimal for the observation of gap solitons. In the following
we will focus on this case. 
We will also consider the effect of a variation
of $N_A$ which allows to test the quasi one-dimensionality
of the gap soliton. 

\section{Stability theory of gap solitons and bright solitons}
A stationary solution $\psi_0$ of the GPE, with chemical potential
$\mu$, is called {\em dynamically stable} if a 
small deviation $\delta\psi$ from $\psi_0$ will not grow
with time. In this case a small perturbation will not cause 
the solution to
evolve into a completely different wavepacket. To study dynamical
instability one can either directly integrate the GPE or solve the
associated Bogoliubov-de Gennes equations (BDGE, see, e.g., 
Ref.~\cite{lewenstein96}). The latter
arise when one writes the wavefunction $\psi({\bf x},t)$ in the form
\begin{equation}
   \psi({\bf x},t) = \exp (-i \mu t/\hbar) (\psi_0({\bf x}) + 
   \delta\psi({\bf x},t))
\label{psiansatz}\end{equation}
and linearizes the GPE in $\delta\psi$.
By making the ansatz of a stationary perturbation,
\begin{equation}
  \delta\psi({\bf x},t) = 
   u({\bf x}) \exp (-i \omega t) - v^*({\bf x})  \exp (i \omega t)\; ,
\label{qpansatz}
\end{equation}
one arrives at the BDGE
\begin{eqnarray}
  \hbar\omega u  &=& {\cal L} u - \kappa \psi_0^2 v \nonumber \\
  -\hbar\omega v &=& {\cal L} v - \kappa (\psi_0^*)^2 u
\label{bdge}
\end{eqnarray}
with ${\cal L} := {\bf p}^2/(2M) + V({\bf x}) -\mu + 2 \kappa |\psi_0|^2$.
A solution $(u,v)$ with eigenvalue $\omega$ corresponds to a
quasi-particle mode. The set of all $\omega$ forms the
quasi-particle spectrum which in general is complex. One can show
\cite{castin01} that if $\omega$ is in the spectrum then so is
$\omega^*$. 
Using Eqn.~(\ref{qpansatz}) it is seen that
the existence of a nonzero imaginary part of one quasi-particle
frequency implies exponential growth of the mode and hence 
dynamical instability of the state $\psi_0$.
To demonstrate that the gap soliton is stable therefore
amounts in showing that the associated quasi-particle spectrum is real.

\section{Numerical methods}
Although the direct numerical integration of the GPE is easy to
implement using the split-step method \cite{deVries86} it has the
disadvantage of not being practical for a 3D study of the gap soliton.
The reason is that the state (\ref{gapsol}) includes two very
different spatial scales: the laser
wavelength $2\pi /k_L$ and the width $\tilde{\textrm{w}}$ of the soliton's
envelope. To cover both scales simultaneously it was necessary to
consider at least 260 periods of the optical potential or 2000
spatial points along the z-axis. Since the number of
total points in 3D is restricted by the capacity of the computer 
a 3D simulation of a gap soliton becomes impractical.
We therefore have applied this method only to 1D and 2D simulations.

To derive the spectrum of quasi-particles around the gap soliton
we followed the method of Ref.~\cite{lewenstein96}
and expanded the modes $(u,v)$ of Eq.~(\ref{bdge}) as well as the 
gap soliton wavefunction $\psi_0$ in a set of basis functions.
For the z-direction we have chosen a number of $n_z$
Bloch wavefunctions which are
eigenstates of the linear Schr\"odinger equation with potential
$V(z) = - V_0 \cos (2 k_L z)$. As transverse modes we used $n_x n_y$
harmonic oscillator eigenstates corresponding to the transverse
trapping potential. This turns Eq.~(\ref{bdge}) into an eigenvalue
problem for a $2n \times 2n$ matrix, where $n := n_x n_y n_z$
is the total number of mode functions. However, before this
eigenvalue problem can be solved, one first has to find the exact
wavefunction of the gap soliton in the given set of basis modes
(in practice a reasonably large subset is sufficient).

Since the analytical solution (\ref{gapsol}) is only approximately correct
a stability analysis will inevitably lead to a complex
quasi-particle spectrum because of the finite difference to the
exact solution. To find the exact solution 
$\psi_{\mbox{{\scriptsize exact}}}$
we have used a
self-consistent field approach (SCF): after expansion of the GPE in the
basis modes Eq.~(\ref{gpe}) is turned into a set of coupled
nonlinear algebraic equations for the expansion coefficients
of $\psi_0$. Using the approximate solution (\ref{gapsol}) as an
ansatz $\psi_{\mbox{{\scriptsize trial}},1}$ we insert it into the GPE
to evaluate the nonlinear terms. The resulting equation,
\begin{equation}
 E \psi({\bf x}) = \left ( \frac{{\bf p}^2}{2M} + V({\bf x}) +
  \kappa |\psi_{\mbox{{\scriptsize trial}}}({\bf x})|^2 \right ) 
  \psi({\bf x})\; ,
\label{gpetrial}
\end{equation}
represents a linear eigenvalue problem for $\psi$ and can easily be
solved using standard numerical methods. We pick that solution
$\psi_{\mbox{{\scriptsize trial}},2}$
out of all eigenstates of Eq.~(\ref{gpetrial}) which has the least
deviation from our previous guess $\psi_{\mbox{{\scriptsize
      trial}},1}$ and iterate this procedure until the
change in the trial wavefunction is below a given value (we used a
relative change of $10^{-14}$ as accuracy goal). The converged
wavefunction may then correspond to the true gap soliton. 

\section{One-dimensional results}
\begin{figure}[hptb]
  \epsfig{file=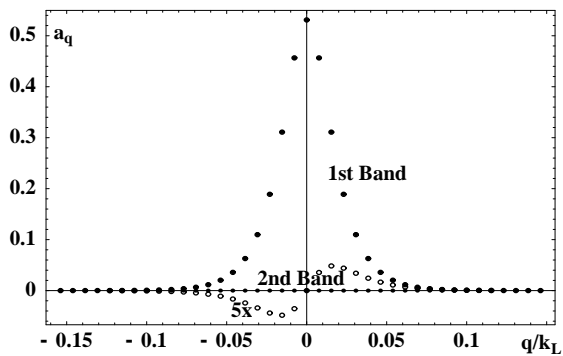, width=7.5cm}
\caption{
The expansion coefficients $a_{q}= \langle\varphi_{q}(z)|\psi(z)\rangle$ 
of the wavefunction $\psi$ in the basis of Bloch functions $\phi_n$ around 
the band edge of the first and second band. 
The circles indicate the initial wavefunction and the dots 
indicate the converged wavefunction. The coefficients of the
first band are nearly identical. The coefficients
of the second band are all close to zero. The coefficients of the
initial wavefunction in the second band have been multiplied by 5
to improve the presentation.
}\label{expcoeffs}
\end{figure}

In one dimension convergence of the SCF algorithm can easily be
achieved. We have used up to 180 Bloch modes to expand the exact
soliton wavefunction which covered the upper half of the lowest
energy band and the lower third of the second Bloch energy band.
However about 40 modes covering the effective-mass region around the
upper band edge were sufficient to get about the
same numerical accuracy. The (real) expansion coefficients
for the gap soliton can be seen in Fig.~\ref{expcoeffs}.
It is interesting
to note that the SCF method essentially amounts to removing that part
of the approximate solution (\ref{gapsol}) which corresponds to
the second energy band. The exact numerical solution therefore
is indeed centered around the upper band edge of the first energy band
where the effective mass is approximately constant.
We have checked whether the final result of the SCF algorithm indeed
describes a soliton by using it as initial condition for the split-step
direct integration of the GPE. It was found that the solution 
does not change its shape for times exceeding 0.2 s.

\begin{figure}[hptb]
\epsfig{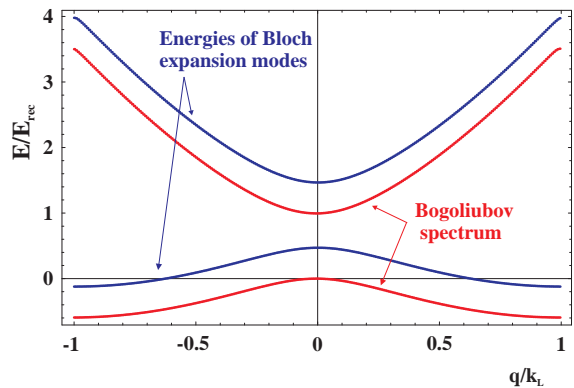}
\caption{Real part of the Bogoliubov spectrum and energies of the
  expansion modes for the first two
  bands in units of the recoil energy $E_{rec}=\hbar^2k_L^2/(2M)$. 
}\label{1dspectrum}
\end{figure}

To verify the stability of the 1D gap soliton we have calculated the
quasi-particle spectrum for up to 350 Bloch modes as basis functions.
The result for the real part of the spectrum can be seen in
Fig.~\ref{1dspectrum} together with the Bloch mode energies. As one
can see the two spectra are very similar apart from a constant shift.
This shift is given by the chemical potential and arises 
because of the corresponding phase factor in ansatz (\ref{psiansatz})
The similarity of the curves is due to the small number of
condensed atoms ($N_A=400$) and the correspondingly small collision effects.

The imaginary part of the quasi-particle spectrum is zero except
for a single mode (and the corresponding mode with complex conjugated
frequency) for which the frequency is purely imaginary. 
The value of the imaginary part for this mode
depends on the degree
of convergence of the SCF wavefunction and on the number of modes
included. For the analytical, non-converged solution (\ref{gapsol})
the decay factor is about $0.02 \times \omega_\perp$, 
where $\omega_\perp$ is the transverse trap frequency.
For a well-converged solution it is always very small,
below $0.001 \times \omega_\perp$. 

To understand the unstable mode better a few general facts about the
quasi-particle spectrum are helpful: for any potential and any
stationary solution of the GPE the mode $(u,v) = (\psi_0,\psi_0^*)$
is a quasi-particle mode with frequency $\omega=0$. This mode is a
Goldstone mode associated with the symmetry of the energy functional
\begin{equation} 
  E = \int \left \{ 
   \psi_0^* \left ( \frac{{\bf p}^2}{2M} + V\right ) \psi_0 +
  \kappa |\psi_0|^4 \right \} d^nx
\label{totenergy}\end{equation}  
with respect to a global phase change $\psi^\prime = \exp (i\alpha)
\psi$. If the external potential $V$ is absent then, for any
stationary solution $\psi_0$ of the GPE, there is a second Goldstone
mode $(u,v) = (\nabla \psi_0,-\nabla \psi_0^*)$. It arises because
of the invariance of $E$ against spatial translations. If the
effective mass approximation would be exact then the gap soliton
would fulfill the same equation as the bright soliton does in free
space. Consequently, it would also possess a translational Goldstone
mode. However, since the periodic optical potential does explicitly
break translational invariance one can expect a shift of the
complex Goldstone frequency away from zero. A second effect that
explicitly breaks translational invariance is the finite number of
Bloch basis functions used in the numerical calculations.
This is equivalent to an optical lattice placed in a box whose
length is a finite multiple of the lattice period. In our case, the
box contained 260 periods. 

\begin{figure}[hptb]
\epsfig{file=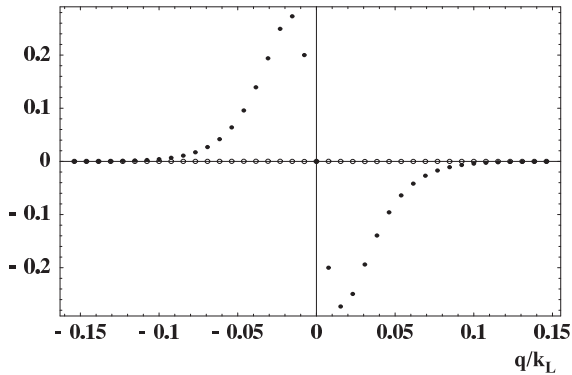, width=7.5cm}
\caption{Expansion coefficients for the modified translational Goldstone
  mode. Shown is the real part (dots) and imaginary part (circles)
  of the function $u(z)$. Numerically it was found that $v(z)\approx -u(z)$.
}
\label{instabmode}
\end{figure}

That the unstable mode indeed corresponds to a modified Goldstone mode
can also be seen by looking at its expansion coefficients 
(Fig.~\ref{instabmode}).
Obviously the shape of $u(q)$ is
approximately given by $q \psi_0(q)$ which would describe the
Goldstone mode if the quasi momentum $q$ is replaced by the real momentum
as it is done in the effective mass approximation.
The question remains whether this tiny instability results from
numerical aberrations, from the finite number of periods, 
or whether it is a real physical effect. To shed some light on this
question we also have performed a stability analysis of the bright
soliton in free space using the same algorithm. We found a similar
behaviour: a Goldstone mode develops an imaginary eigenvalue, but
this time it is the mode associated with the phase transformation.
Since one can prove that this mode has zero frequency we conclude
that the tiny imaginary eigenvalue for Goldstone modes is a
spurious numerical effect and the gap soliton is dynamically stable
in one dimension.

\section{Results in two and three dimensions}
Having examined the stability of the 1D gap soliton it is of 
interest whether it will remain stable in a quasi one-dimensional
situation. The condition for the latter is usually formulated
as follows: the interaction energy, which leads to a coupling between
different modes of the corresponding linear Schr\"odinger equation,
should be much smaller than the excitation energy of the transverse
trapping potential. If this is fulfilled a BEC in its ground state
will effectively behave like a one-dimensional quantum gas
since transverse excitations are off-resonant and therefore suppressed.

\begin{figure}[hptb]
\epsfig{file=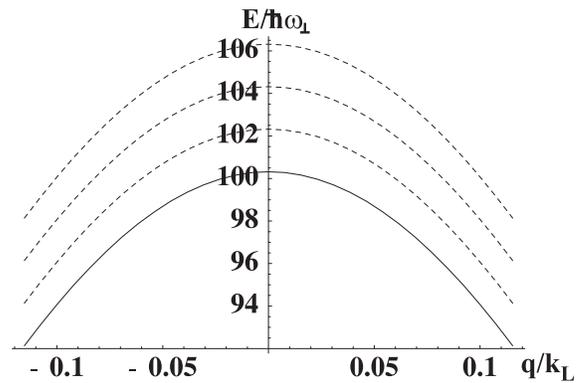, width=7.5cm}
\caption{Energy eigenvalues around the upper
band edge for noninteracting atoms in an optical
lattice and with a transverse trapping potential. Due to resonances
between longitudinal and transverse excitations the gap soliton
will be unstable against transverse decay. The physical
parameters are given in the text.}
\label{freeenergies}
\end{figure}

This is not the case for a gap soliton, however. The reason can be seen
by looking at Fig.~\ref{freeenergies} which displays the mode
energies of noninteracting atoms in the optical lattice and with
a tight harmonic transverse confinement around the upper band edge.
The solid line displays the energy of atoms in the transverse
ground state and with longitudinal quasi momentum $q$ in the
lowest energy band. Each dashed line corresponds to transversally
excited atoms with a transverse energy of $2\hbar \omega_\perp$
to $6\hbar\omega_\perp$, respectively. It is important to
observe that there are resonances between
transversally unexcited atoms with $q=0$ and transversally excited
atoms with $q\neq 0$.
Since the gap soliton is a superposition of Bloch modes around
the upper band edge ($q=0$), these resonances have the consequence
that even for tight transverse confinement a true gap soliton
does not exist. This situation is qualitatively the same in two and in three
dimensions since in both cases the free energy levels are given
by those of Fig.~\ref{freeenergies}. Since in two
dimensions the transverse trapping potential is one-dimensional
the multiplicity of the energy levels is always one.
This is different in three dimensions where a
transverse  excitation energy of $n\hbar \omega_\perp$ has an $n+1$ fold
degeneracy. The number of resonant states is therefore larger
than in two dimensions.

Although a true gap soliton does not exist it is of interest to
examine a {\em quasi gap soliton} of the form
\begin{equation} 
 \psi_{\mbox{{\scriptsize quasi}}} =  \psi_{\mbox{{\scriptsize gap}}}(z,t)
    \varphi_0 (x,y)\; ,
\label{quasi} 
\end{equation} 
with $\varphi_0$ denoting the transverse ground state.
It should be possible to produce a state like 
$ \psi_{\mbox{{\scriptsize quasi}}}$ using dispersion management
\cite{treutlein02}.
Though the quasi gap soliton is not a true stationary solution of the GPE, 
it may be sufficiently stable to allow for experimental observation.
A signature of a quasi gap soliton would be a strongly suppressed
dispersion of the wavepacket along the z axis.
To analyze the time scale on which this state decays we first note
that the transverse ground state
has even parity and  because of parity conservation 
can only couple to even excited
levels $ 2n\hbar\omega_\perp$. This is the reason why we omitted
odd transverse excitations in Fig.~\ref{freeenergies}.

\begin{figure}[hptb]
\epsfig{file=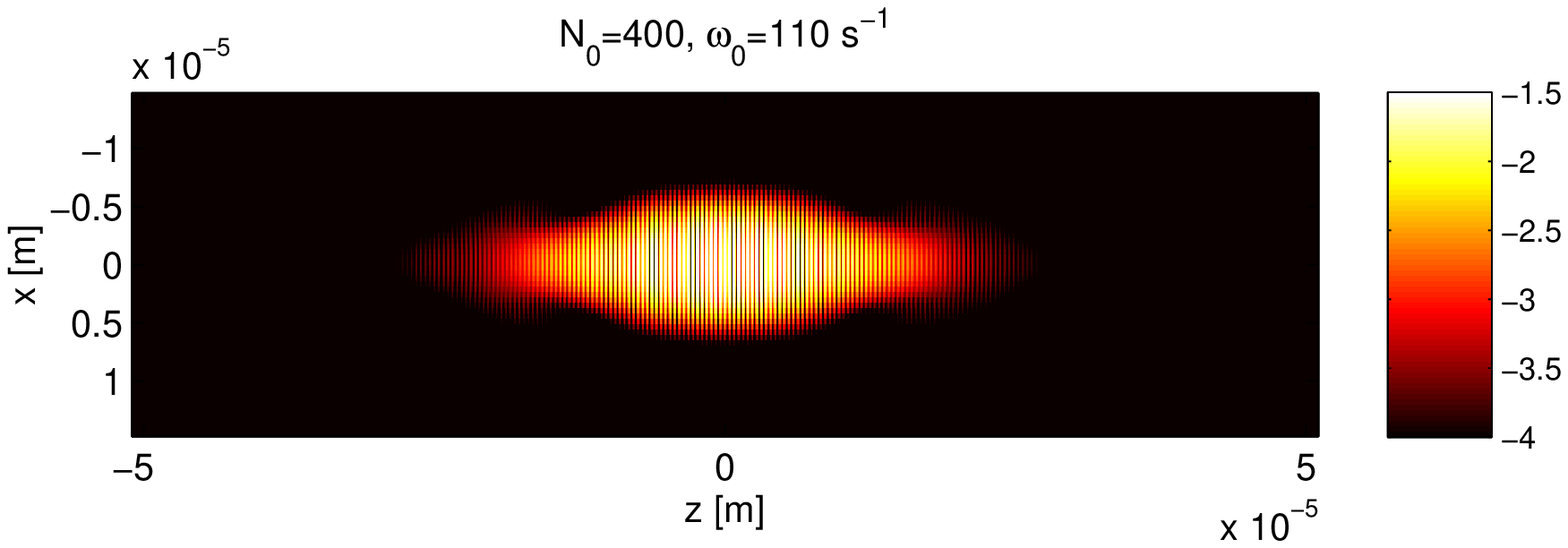,width=8cm}\put(-225,72){A)}\vspace{0.1cm}
\epsfig{file=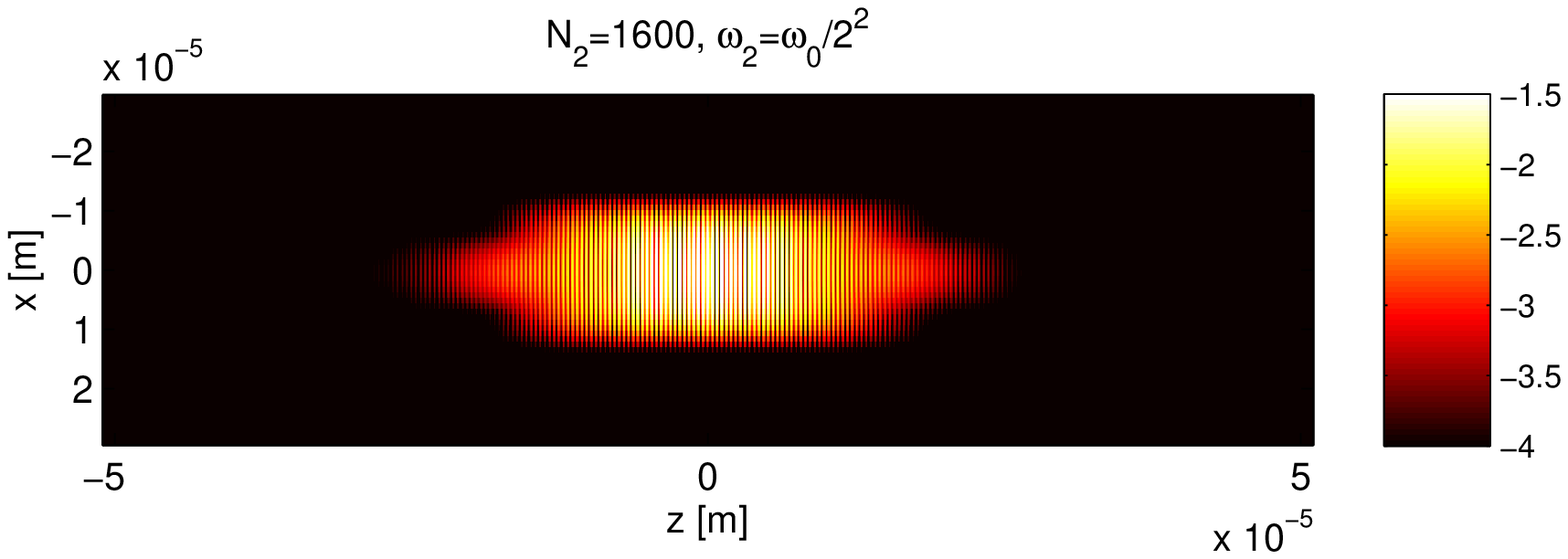, width=8cm}\put(-225,72){B)}\vspace{0.1cm}
\epsfig{file=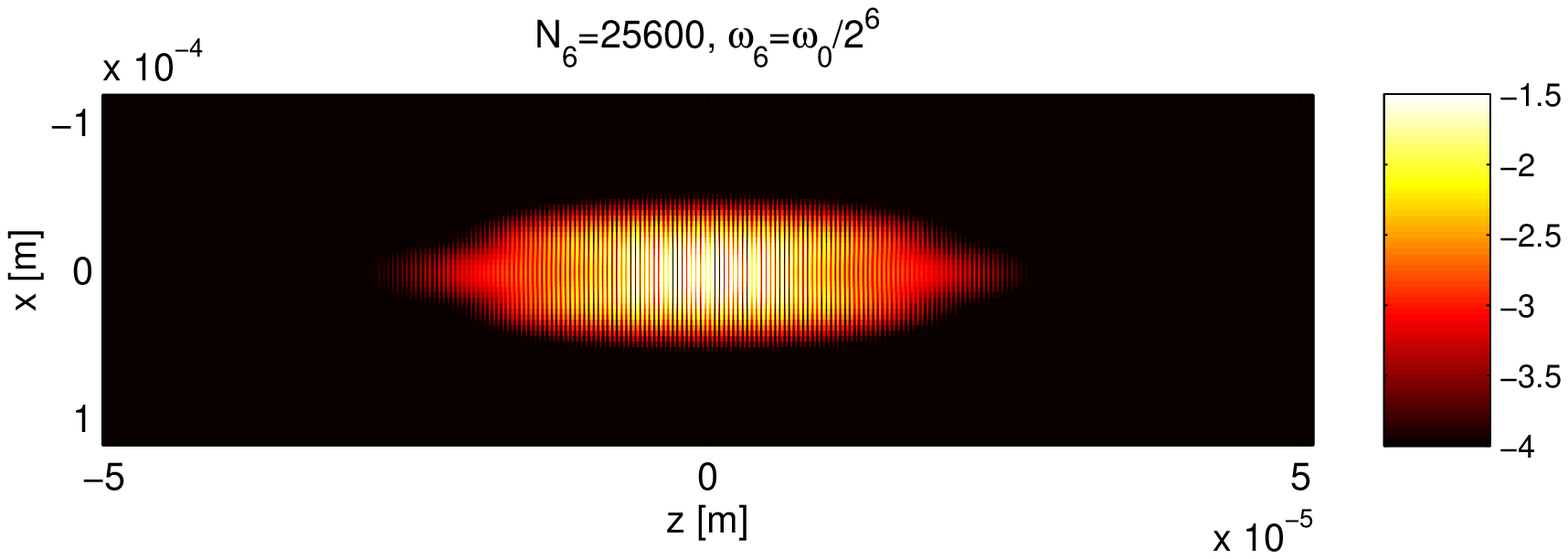,width=8cm}\put(-225,72){C)} 
\caption{Density of the wavefunction 
$\log_{10}|\psi|^2$ at $t=23ms$ for A) 400 atoms, B) 1600 atoms,
and C) 25600 atoms in the BEC. See text for details.
 }\label{2dgpepics1}\end{figure}

\begin{figure}[hptb]
\centerline{\epsfig{file=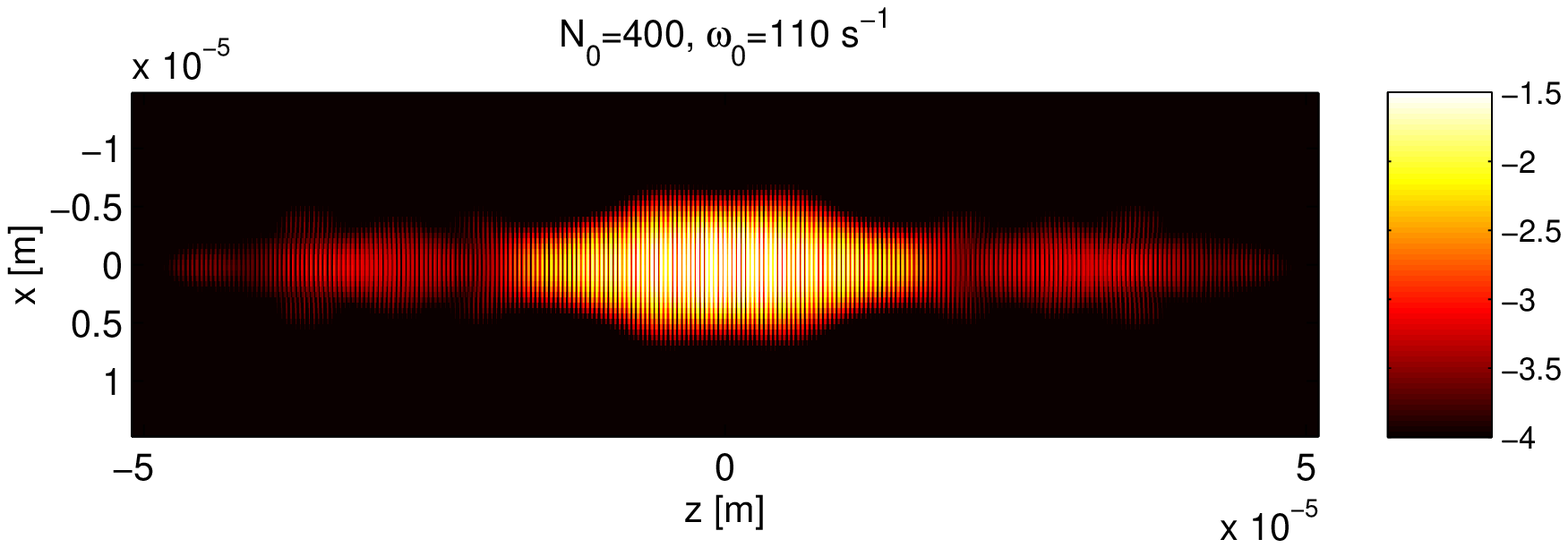, width=8cm}\put(-225,72){A)}}\vspace{0.1cm}
\centerline{\epsfig{file=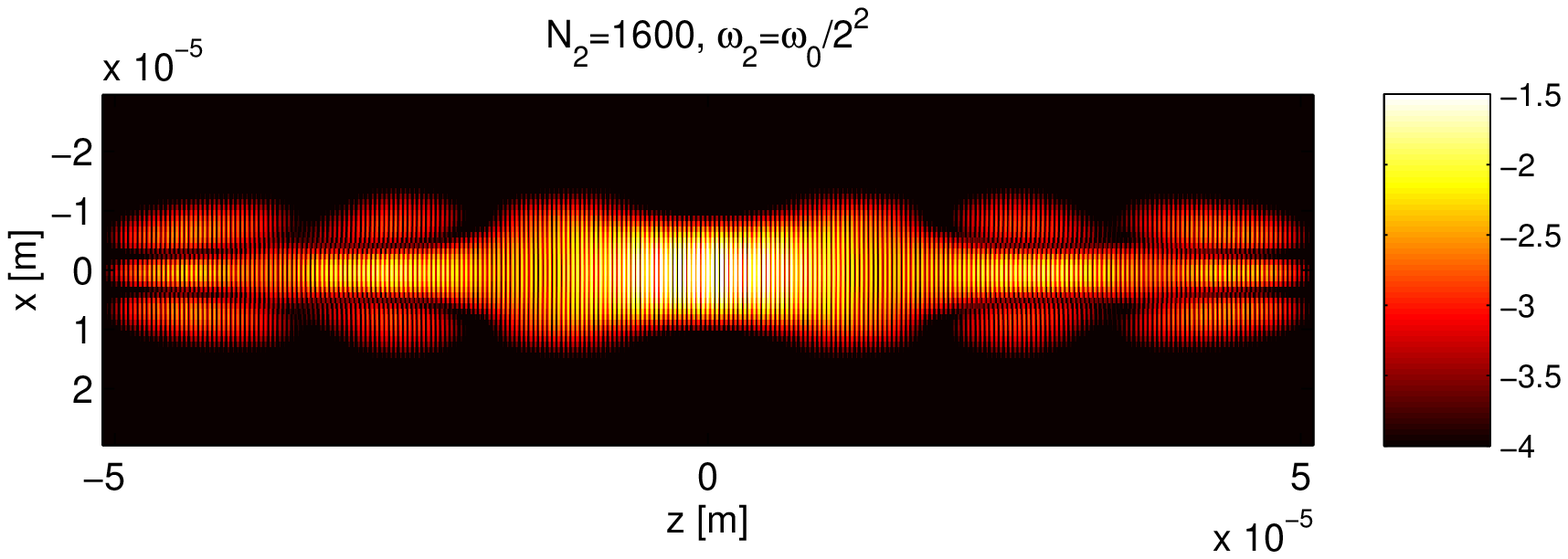, width=8cm}\put(-225,72){B)}}\vspace{0.1cm}
\centerline{\epsfig{file=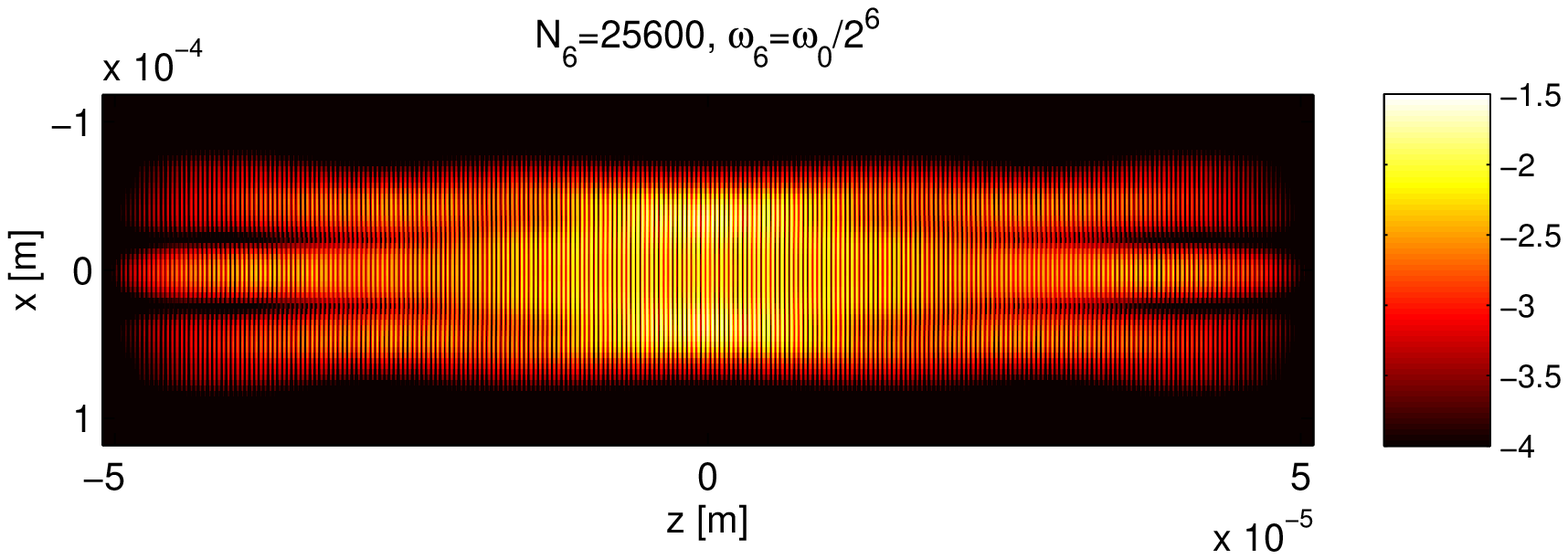, width=8cm}\put(-225,72){C)}}
\caption{Density of the wavefunction 
$\log_{10}|\psi|^2$  at $t=0.23s$.  for A) 400 atoms, B) 1600 atoms,
and C) 25600 atoms in the BEC. See text for details.}\label{2dgpepics2}
\end{figure}

We have used the state (\ref{quasi}) as initial condition and studied
numerically the time evolution of it in two dimensions.  To study the
influence of the transverse confinement we have considered three BECs
with 400, 1600, and 25600 atoms in Figs.~\ref{2dgpepics1} A, B, and C,
respectively. The number of atoms  $N_A$ and the transverse
confinement frequency $\omega_\perp$ have been simultaneously varied
keeping the product $\omega_\perp N_A$ constant. Consequently the
interaction energy in Eq.~(\ref{totenergy}) is kept constant, 
since it is proportional to $(\omega_\perp N_A)^2$. 
This also ensures that the first-order soliton condition is fulfilled.
The result of the numerical time evolution after 23 ms is shown in Fig.~\ref{2dgpepics1}. While some
excitations are observable, the state still looks very much like a
quasi gap soliton. This situation changes after 0.23 s
(Fig.~\ref{2dgpepics2}). Here one
can see strong excitations which are growing with decreasing transverse excitation frequency. This is a reasonable result since the ratio of the interaction energy and the transverse excitation energy is greater than one for Fig.~\ref{2dgpepics2} B and C hence the nonlinear coupling is strong enough to excite the transverse modes. The figure suggests that in particular the transverse modes with
$2\hbar\omega_\perp$ energy are strongly excited because
for a fixed value of $z$ there are three density maxima in the x-direction.
This is in agreement
with the qualitative predictions which we have made above. 

To verify this result we also have calculated the quasi-particle 
spectrum of the state
(\ref{quasi}) in two and three dimensions. In principle, since
the quasi gap soliton is not a stationary state, the spectrum
is not enough to predict the evolution of it accurately
and a more sophisticated approach is needed \cite{castin97}.
However, to gain a qualitative understanding of the time scale
on which $\psi_{\mbox{{\scriptsize quasi}}}$ decays the 
imaginary part of the spectrum is sufficient. 

\begin{figure}[hptb]
\centerline{\epsfig{file=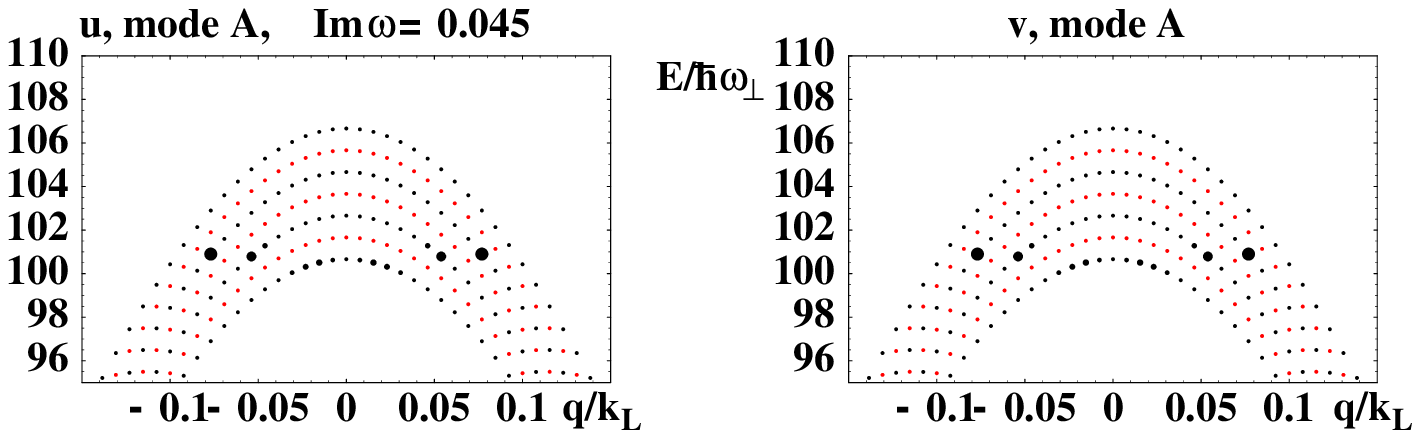, width=9cm}}\vspace{0.1cm}
\centerline{\epsfig{file=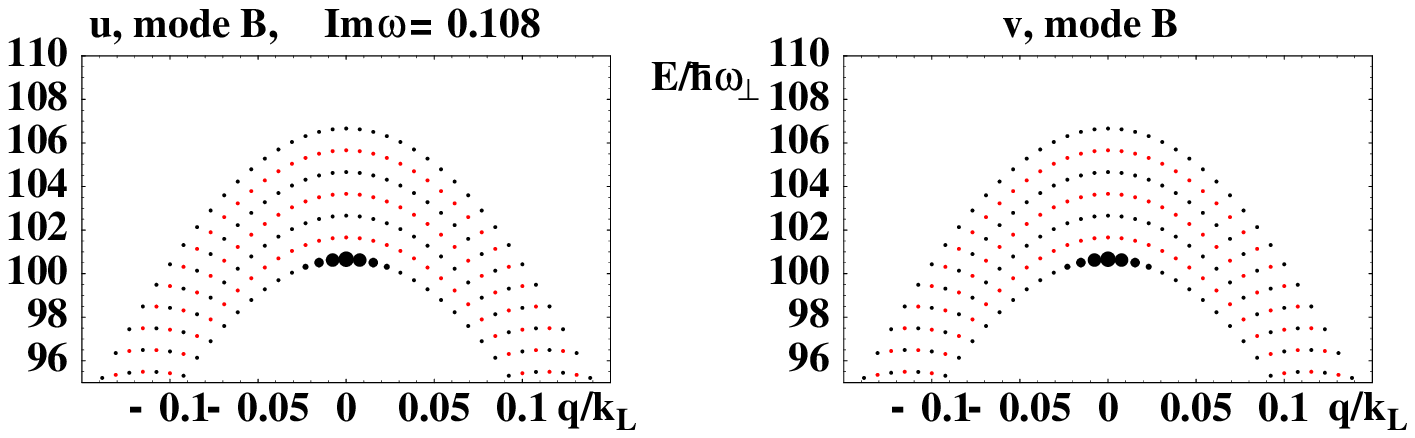, width=9cm}}\vspace{0.1cm}
\centerline{\epsfig{file=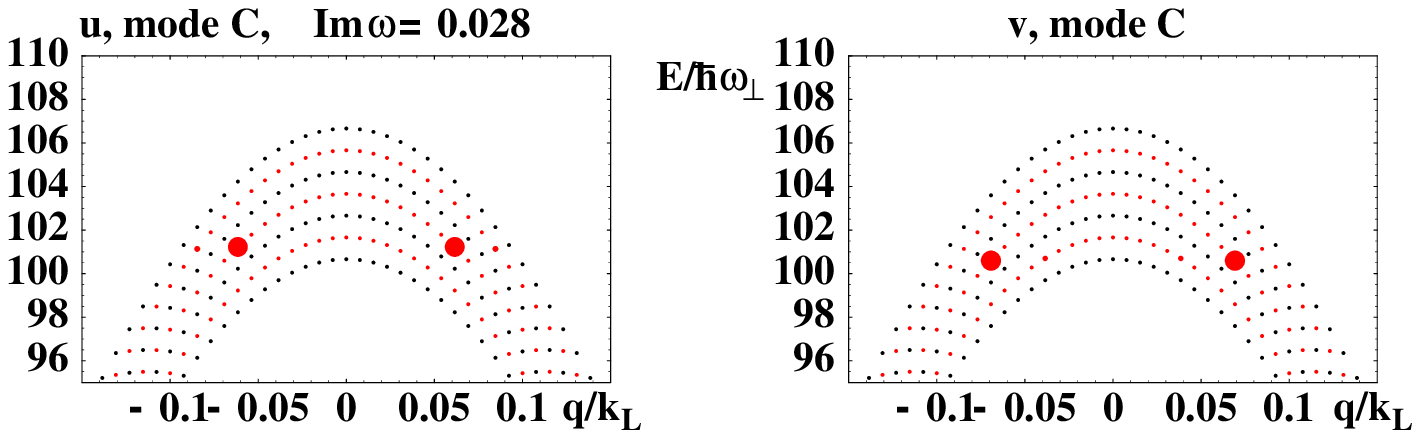, width=9cm}}
\centerline{\epsfig{file=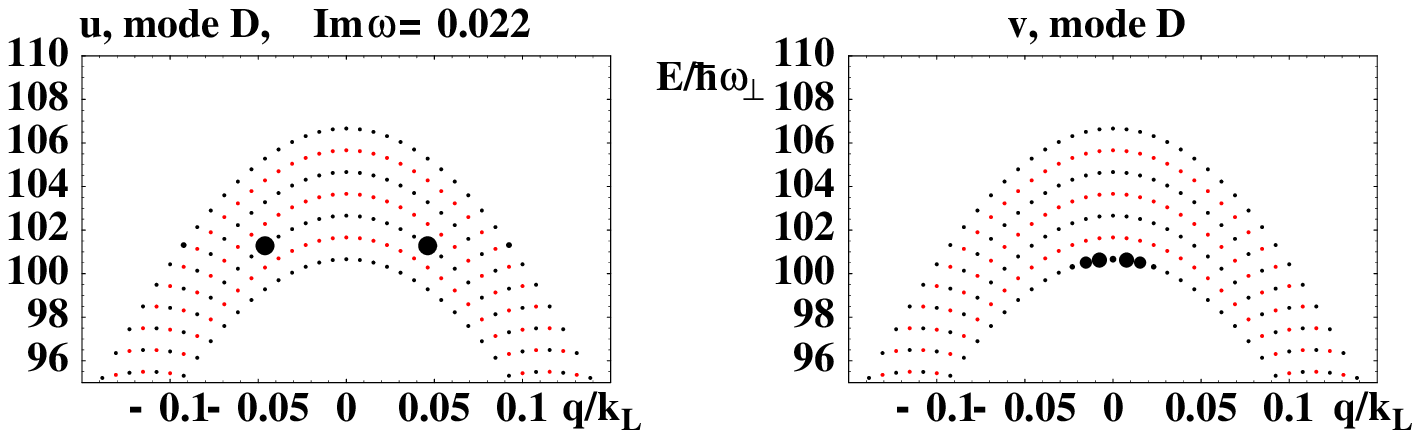, width=9cm}}
\caption{Selected unstable Bogoliubov modes for
a 2D quasi gap soliton. Shown is the energy $E$ (in units
of $\hbar \omega_\perp$) of the basis functions as a
function of the quasi momentum $q$ (in units of $k_L$).
The lowest parabola corresponds to the transverse ground state.
The other parabolas describe a transverse excitation of
$n \hbar \omega_\perp, n=1,2,\ldots$.
The thickness of the dots corresponds to
the modulus of the u or v coefficients with respect to the
corresponding basis function. The smallest dots correspond
to coefficients of modulus between 0 and 0.1. The imaginary part Im($\omega$)
of the unstable modes is given in units of the transverse
trap frequency $\omega_\perp = 110$ s$^{-1}$.
}\label{2dmodes}
\end{figure}

In 2D we used up to 100 Bloch wavefunctions and up to 19
one-dimensional eigenstates of the transverse harmonic trap as basis modes.
The quasi gap soliton was expanded using 51 Bloch states. It turned out that 
there are generally quite many unstable modes, but only few of them do
have a considerable overlap with the collective wavefunction.
The number of unstable modes depends on the number of basis
states used to expand the BDGE since the number of resonant transversally
excited states is growing. However, it turned out that this basis
dependence does only affect modes with a small instability.
Some examples are displayed in Fig.~\ref{2dmodes}.
Mode \ref{2dmodes}A
corresponds very well to the anticipated resonant excitation
of transverse modes. It has a non-zero overlap with the quasi
gap soliton and otherwise only populates even transversely excited
basis modes. Correspondingly its instability is rather large;
only the instability of mode \ref{2dmodes}B, 
which roughly describes the phase Goldstone mode, decays faster.
Mode  \ref{2dmodes}B is unstable because, as described above,
the quasi gap soliton is not a stationary solution of the GPE.
However, due to a coupling to transversally excited states
the decay rate is strongly enhanced as compared to the non-converged
one-dimensional quasi gap soliton
(the modulus of the transversally excited mode coefficients 
is less than 0.1 and is therefore not visible in Fig.~\ref{2dmodes}).
Mode  \ref{2dmodes}C is typical for the many unstable modes which have no
overlap with the quasi gap soliton. Therefore, unless transversal
excitations are created during the experimental preparation of
the quasi gap soliton, they do not contribute to the decay of it.
These modes appear because the collective wavefunction provides
a linear coupling term between transversally excited states which
are in resonance with each other. Such modes can exist for even and
odd transverse excitation levels without violating parity.
Finally, mode  \ref{2dmodes}D describes a somewhat off-resonant
coupling between the transverse ground state and transversally excited
states. Because of its off-resonant nature its decay rate
Im($\omega$) is considerably smaller than for the resonant mode
 \ref{2dmodes}A.

To analyse the dynamical instability of a 3D quasi gap soliton we
used again Bloch wavefunctions for the longitudinal expansion
of the Bogoliubov modes. For the two transversal directions
we have chosen a basis of states which are both eigenstates
of the Hamiltonian and the angular momentum operator $L_z$.
These states can be constructed by using the creation operators
$c_\pm := (a_x^\dagger \pm i a_y^\dagger)/\sqrt{2}$
\cite{marzlin98}. The basis states are then given by
\begin{equation}
  |n,m\rangle := \frac{ (c_+^\dagger)^{(n-m)/2}
   (c_-^\dagger)^{(n+m)/2} }{ \sqrt{((n+m)/2)! ((n-m)/2)!}}|0\rangle\; .
\label{harmang}\end{equation}
The energy of these states is given by $n\hbar \omega_\perp,
n=0,1,2,\ldots$ and their angular momentum by 
$m\hbar, \; m=-n, -(n-2), \ldots , n$. We used again 51 Bloch states
to expand the Bogoliubov modes along the z-axis and up to 30
transverse modes as given in Eq.~(\ref{harmang}).

\begin{figure}[hptb]
\centerline{\epsfig{file=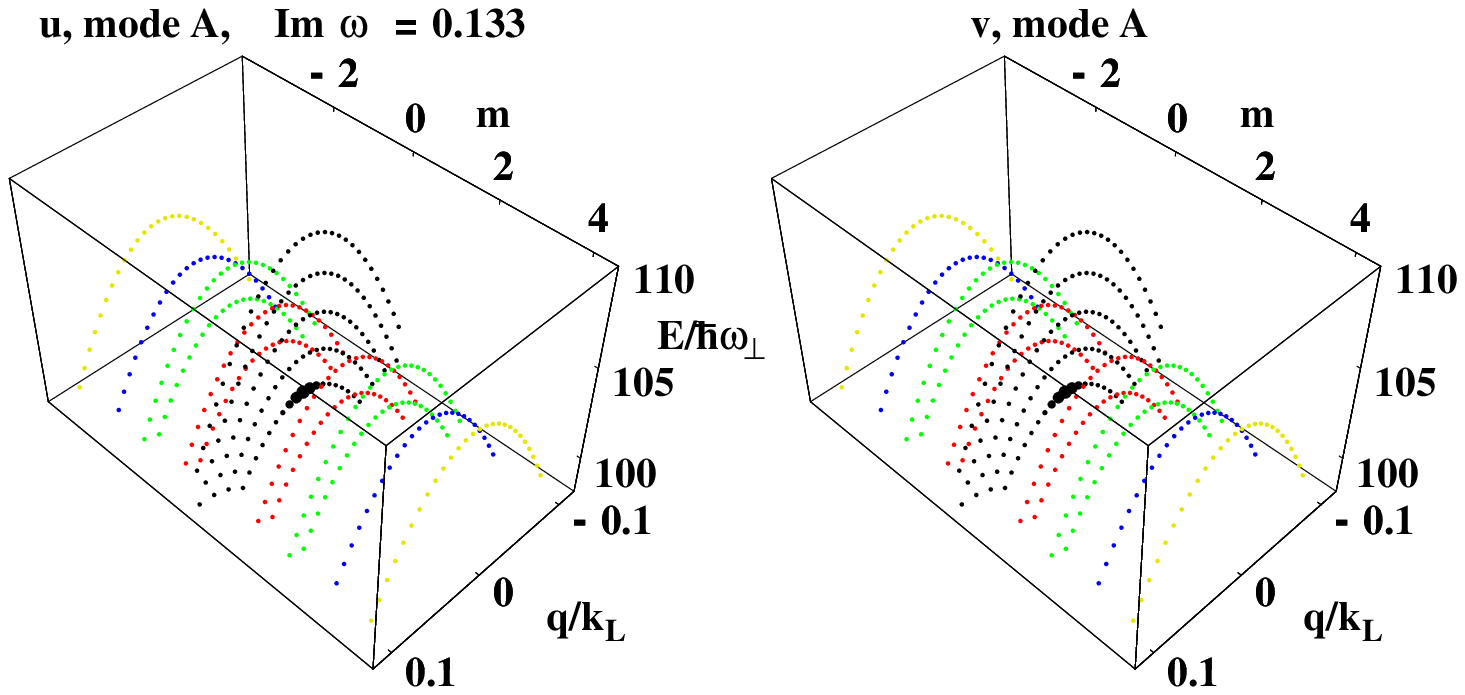, width=9cm}}\vspace{0.1cm}
\centerline{\epsfig{file=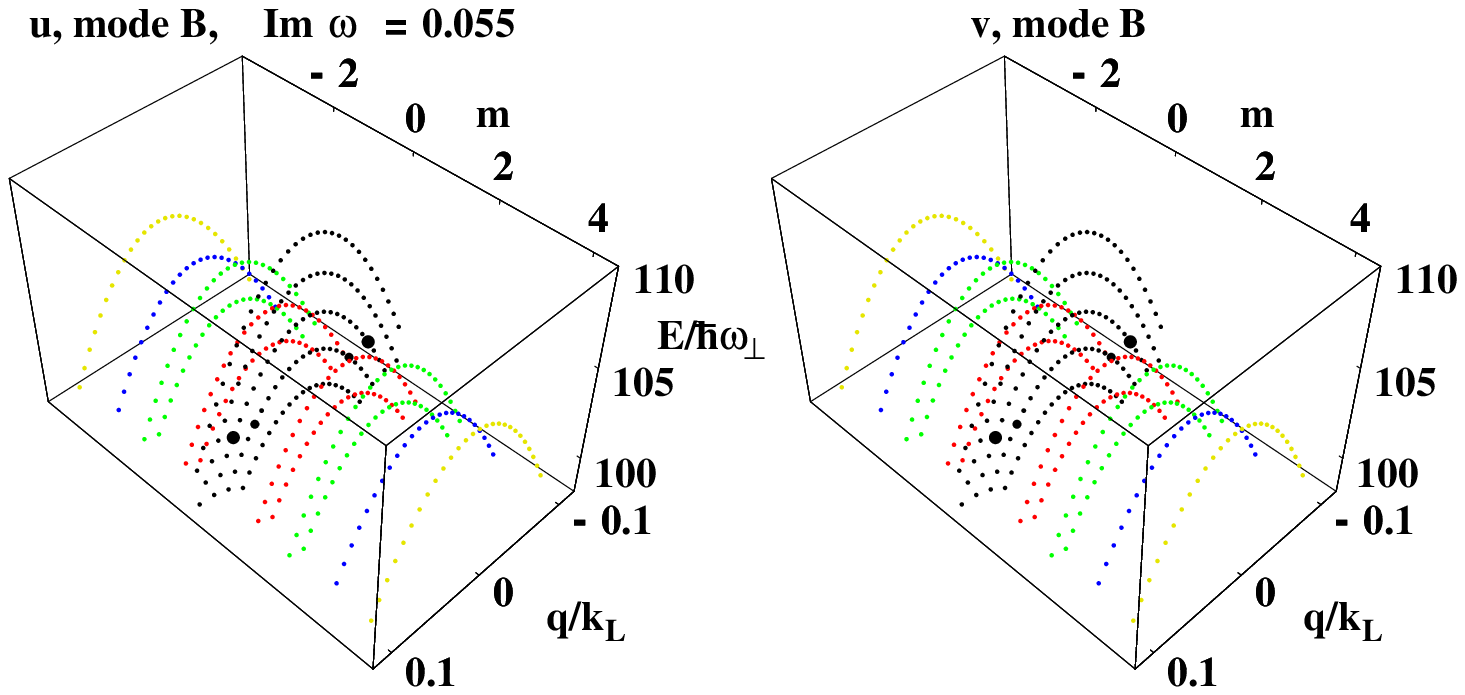, width=9cm}}\vspace{0.1cm}
\centerline{\epsfig{file=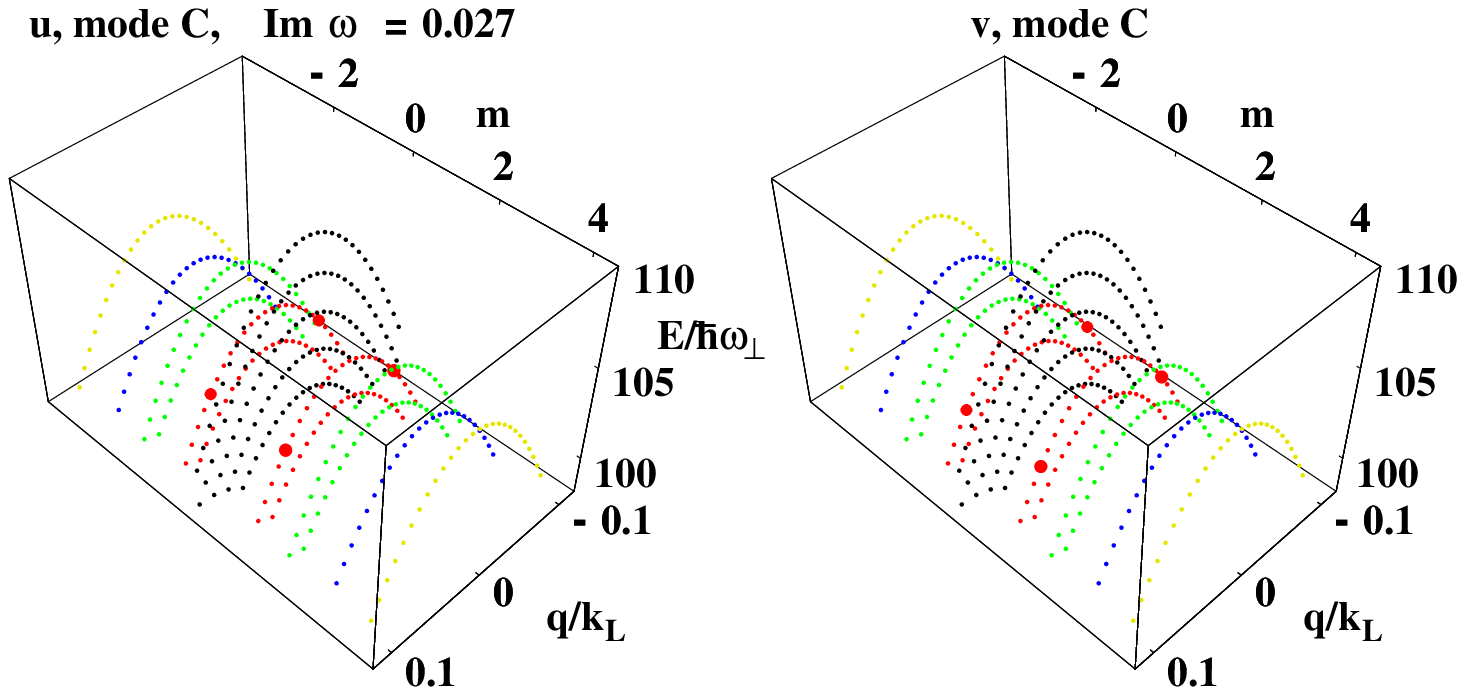, width=9cm}}
\caption{Selected unstable Bogoliubov modes for
a 3D quasi gap soliton. The units are as in Fig.~\ref{2dmodes}, but
in addition to the basis energy $E$ and quasi momentum $q$ a third
quantum number $m$ is introduced which is associated with the
axial angular momentum $L_z = m \hbar$ of the basis functions.
}\label{3dmodes}
\end{figure}

We found again a large number of unstable modes which 
spuriously depends on the number of transverse basis states.
However, similarly to the 2D case, modes with an
instability Im$(\omega) > 0.02 \omega_\perp$ do not
depend significantly on the number of basis states.
Some of these unstable Bogoliubov modes in 3D are shown in Fig.~\ref{3dmodes}.
Mode \ref{3dmodes}A is again roughly the phase Goldstone mode plus some
small transversal excitations not visible in the figure. Due to
the larger number of nearly resonant basis states the decay rate is even larger
than in the 2D case. Mode \ref{3dmodes}B is again a superposition of resonant
states and, as in 2D, has the second largest decay rate.
These two states as well as most of the other unstable modes
are solely composed of states with zero angular momentum,
i.e., states which are symmetric in the x-y plane. They are therefore
very similar to the 2D case.
A new kind of
instability in 3D occurs in Mode \ref{3dmodes}C which has no overlap with the
quasi gap soliton. It is composed out of states with $m=\pm 1$
and thus describes the decay or increase of rotating states
linearly coupled by the collective wavefunction. We found 80
unstable modes of this kind.

\section{Conclusion}
We have examined the dynamical instability of gap solitons in a BEC
in one, two, and three spatial dimensions under the condition of
tight transversal confinement. Using different methods we found
that a truly one-dimensional gap soliton is stable. In higher
dimensions transversal excitations which are resonant with
the upper band edge forbid the existence of a real gap soliton.
However, a quasi gap soliton may be experimentally prepared which
behaves like the 1D gap soliton for a time smaller than 
the smallest decay time of one of the unstable Bogoliubov
modes. In 3D we numerically found this time to be in the
order of $1/(0.133 \times \omega_\perp)$ which is the decay time
of mode A in Fig.~\ref{3dmodes}. For a transverse trap frequency
of $\omega_\perp = 110$ s$^{-1}$ we expect the quasi
gap soliton to be sufficiently stable for about 70 ms.
This should be long enough for experimental observation.

Acknowledgement:
We are indebted to Klaus M{\o}lmer and Gora Shlyapnikov for very
valuable discussions and J\"urgen Audretsch for kind hospitality.
This work was supported by the Optikzentrum Konstanz and the
Forschergruppe Quantengase. M.K.O. is supported by the Emmy Noether
Programm of the Deutsche Forschungsgemeinschaft.
K.M.H. would like to thank the following fonds for financial support: 
Familien Hede Nielsens Fond, Krista og Viggo Petersens Fond, 
Observator mag.~scient Julie Marie Vinter Hansens Fond, 
Otto Bruuns Fond Nr.~2, Etatsraad 
C.G. Filtenborg og hustru Marie Filtenborgs studielegat, Rudolph Als Fondet,
Frimodt-Heineke Fonden, Sokrates/Erasmus studierejselegat
and Professor Jens Lindhards Forskningslegat.

\end{document}